\documentclass[twocolumn,epjc3]{svjour3}          

\RequirePackage[T1]{fontenc}

\usepackage{amsmath,amssymb,booktabs}
\usepackage{microtype}

\RequirePackage{graphicx}
\RequirePackage{mathptmx}      
\RequirePackage{flushend}
\RequirePackage[numbers,sort&compress]{natbib}
\RequirePackage[colorlinks,citecolor=blue,urlcolor=blue,linkcolor=blue,breaklinks=true]{hyperref}
\renewcommand{\thesection}{\Alph{section}}
\newcommand{\Tr}{\operatorname{Tr}}
\newcommand{\dd}{\mathrm{d}}
\newcommand{\ii}{\mathrm{i}}
\newcommand{\ord}{\mathcal{O}}
\newcommand{\diag}{\operatorname{diag}}
\newcommand{\id}{\mathbb{I}}
\renewcommand{\thesection}{\arabic{section}}
\journalname{Eur. Phys. J. C}

\begin{document}

\title{Efficiency enhancement in a two-detector Unruh quantum Otto engine}

\author{Yongjie Pan\thanksref{addr1,addr2}
        \and
        Baocheng Zhang\thanksref{e1,addr3}
        \and
        Qingyu Cai\thanksref{e2,addr1,addr2} 
}

\thankstext{e1}{e-mail: zhangbaocheng@cug.edu.cn}
\thankstext{e2}{e-mail: qycai@hainanu.edu.cn}

\institute{Center for Theoretical Physics, Hainan University, Haikou 570228, China\label{addr1}
          \and
          School of Physics and Optoelectronic Engineering, Hainan University, Haikou 570228, China\label{addr2}
          \and
          School of Mathematics and Physics, China University of Geosciences, Wuhan 430074, China\label{addr3}
}

\date{Received: date / Accepted: date}

\maketitle

\abstract{We investigate efficiency enhancement in an Unruh quantum Otto engine (UQOE) consisting of two initially uncorrelated Unruh–DeWitt (UDW) detectors. The detectors undergo uniform acceleration with different magnitudes and couple locally to a massless scalar field prepared in the Minkowski vacuum. We choose a diagonal free Hamiltonian for the UQOE and define its work and heat with respect to the proper time of one of the detectors. Consequently, the coherence terms generated through the interaction with the scalar field do not directly contribute to the work and heat of the UQOE. Instead, the efficiency enhancement arises from the relative proper-time weighting of the detectors' local population changes. We identify operating points that satisfy the conditions of initial-state passivity, cycle closure, and energy conservation. The efficiency is higher than that of a conventional UQOE, and the state of the UQOE after the heat absorption stage is separable. We further investigate the influence of detector separation on cycle closure and efficiency. Although the local population responses are independent of separation at fixed control parameters, the nonlocal coherence terms depend on separation and impose additional constraints on state cycle closure. Restoring closure generally requires retuning the cycle parameters, which can indirectly modify the achievable efficiency enhancement. Our results demonstrate that, within the second-order perturbative framework and the adopted energy convention, entanglement generation is not necessary for efficiency enhancement. They also clarify the distinct roles of local population changes and nonlocal coherences in the thermodynamic operation of a two-detector UQOE. 
} 
%



\section{Introduction}\label{sec:intro}
Quantum heat engines provide a framework for investigating energy conversion and thermodynamic processes at the microscopic scale. Among various quantum thermodynamic cycles, the quantum Otto cycle consists of two adiabatic strokes, during which the energy gap of the working medium is varied, and two isochoric strokes, during which energy is exchanged with the environment at a fixed energy gap \cite{kieu2004second,quan2007quantum}. For a conventional single-detector quantum Otto engine, the efficiency takes the simple form $\eta_0=1-\omega_1/\omega_2$, where $\omega_1$ and $\omega_2$ are the energy gaps during the cold and hot isochoric strokes, respectively, provided that the cycle closes and the heat and work satisfy the required sign conditions. This expression provides a useful reference for examining the physical mechanisms underlying efficiency enhancement. Previous studies have shown that quantum correlations, including entanglement, can affect the thermodynamic performance of quantum heat engines and may enhance their efficiency under suitable conditions ~\cite{zhang2007four,zhang2008entangled,wang2009thermal}. However, observing an enhanced efficiency in an entangled system does not, by itself, establish entanglement as the physical origin of the enhancement. Identifying the mechanism responsible for such an improvement therefore requires a careful examination of the contributions of quantum correlations to energy exchange and cyclic operation.

The interplay between relativity and quantum thermodynamics has motivated the study of quantum heat engines operating in relativistic settings~\cite{Arias2017Unruh,Gray2018Scalar,Xu2019Unruh}. A particularly relevant phenomenon is the Unruh effect~\cite{unruh1976notes,unruh1984happens,crispino2008unruh}, which establishes a connection between uniformly accelerated motion and the thermal properties of quantum fields. According to the Unruh effect, a uniformly accelerated observer in Minkowski spacetime associates a temperature proportional to the proper acceleration with the Minkowski vacuum~\cite{unruh1976notes,unruh1984happens,crispino2008unruh}. The Unruh–DeWitt (UDW) detector~\cite{dewitt1979quantum}, modeled as a localized quantum system coupled to a field, provides a useful framework for investigating this phenomenon [12]. In particular, a two-level UDW detector can serve as the working medium of an Unruh quantum Otto engine (UQOE), in which accelerated detector–field interactions play the role of the isochoric strokes~\cite{Arias2017Unruh,Gray2018Scalar}. Previous studies have explored the thermodynamic behavior of such engines, including the effects of field statistics and level degeneracy~\cite{Satz2007Switching,Liu2026odq}. In conventional single-detector UQOE models, the detector response determines the population changes and the amount of work extracted, while the efficiency retains its standard form $\eta_0$. It should be emphasized, however, that a finite-time detector–field interaction does not generally lead to complete thermalization. The detector response depends on the switching function and the trajectory, and the corresponding heat and work must therefore be evaluated from the actual state changes rather than inferred solely from the Unruh temperature~\cite{Satz2007Switching,Liu2026odq}.

An important extension of the UQOE model involves two detectors undergoing uniform acceleration with different magnitudes. In this case, their proper times must be related to define the energy of the combined system with respect to a common time parameter~ \cite{Rodriguez2016Finite,Henderson2020Quantum,barman2021radiative,mukherjee2023fulling,bhattacharya2023extraction}. Kane and Majhi investigated an initially entangled two-detector UQOE and obtained an acceleration-dependent efficiency by incorporating the relative proper-time factor into the thermodynamic description  \cite{Kane2021Entangled}. Subsequently, Barman and Majhi included time-ordering effects and examined the positive-work and cyclicity conditions for specified entangled initial states \cite{Barman2021Constructing}. These studies demonstrated that two-detector UQOEs can exhibit efficiency enhancement compared with the conventional Otto value. Nevertheless, an important question remains: What is the physical mechanism responsible for this enhancement, and does it necessarily require entanglement between the detectors? This question is particularly relevant because two locally coupled UDW detectors can develop quantum correlations through their interactions with a common field, even in the absence of direct detector–detector coupling \cite{reznik2003entanglement,reznik2005violating,pozas2015harvesting,martin2016spacetime}. Depending on the spacetime arrangement of the interaction regions, these correlations may arise from vacuum correlations and causal communication~\cite{tjoa2021entanglement,maeso2022entanglement}. Since the Gaussian switching functions considered in the present work have noncompact support, we refer to the resulting process as entanglement generation rather than strictly spacelike-separated entanglement harvesting.

In this work, we investigate a two-detector UQOE with an initially uncorrelated product state, in contrast to the initially entangled configurations considered in Refs.~\cite{Kane2021Entangled,Barman2021Constructing}. The two detectors undergo uniform acceleration with different magnitudes in the same direction and interact locally with a massless scalar field in the Minkowski vacuum. We describe their combined dynamics perturbatively to second order in the detector–field coupling, retaining both the diagonal population changes and the complex off-diagonal coherence terms of the reduced density matrix. The free Hamiltonian of the two-detector working medium is taken to be diagonal, and heat and work are defined with respect to the proper time of detector $A$. This choice introduces a relative proper-time weight into the contribution of detector $B$ to the internal energy. We use the complete state evolution to derive the conditions for state cycle closure and impose energy balance and the appropriate heat–work sign conditions to identify admissible operating points. Unlike conditions derived for initially entangled configurations \cite{Kane2021Entangled,Barman2021Constructing}, the closure constraints must be evaluated for the product-state preparation adopted here.

Our analysis shows that the enhancement of the efficiency is governed by the local population changes and their relative proper-time weighting. For the diagonal free Hamiltonian considered here, the energy expectation value is invariant under dephasing in the product energy basis. Consequently, the field-induced coherences do not contribute directly to the absorbed heat or the work associated with energy-gap modulation. The resulting efficiency enhancement instead arises because the population change of detector $B$ enters the absorbed heat with a relative proper-time weight different from that appearing in the work. In particular, when detector $A$ has a smaller acceleration than detector $B$, a positive population change of detector $B$ leads to an efficiency above the conventional Otto value, provided that the absorbed heat is positive. The existence of a thermodynamically admissible cycle further requires state closure, energy balance, and positive work output. We construct numerical families of operating points satisfying these conditions and find that the efficiency can exceed the conventional Otto value even when the post-hot-stroke state is separable. This establishes that entanglement generation is not a necessary ingredient of the efficiency enhancement mechanism identified within the present framework.

Although the generated coherences do not contribute directly to heat or gap work, they remain relevant to the existence of a closed thermodynamic cycle. State closure requires the off-diagonal elements of the reduced density matrix, as well as its populations, to return to their initial values. This distinction becomes particularly important when considering the separation between the detectors. At fixed accelerations, energy gaps, switching parameters, and initial populations, the local population responses are independent of the transverse detector separation, whereas the nonlocal coherence terms generally depend on it. Consequently, varying the separation can violate the state closure conditions without directly changing the population-dependent heat and work expressions. Restoring closure requires readjusting other control parameters, which may indirectly affect the achievable efficiency. Our analysis therefore distinguishes the local population responses governing energy exchange from the nonlocal coherence constraints governing state cycle closure .

Section~\ref{sec:model} presents the interaction model and the quantum Otto cycle. Section~\ref{sec:energy} presents the expressions for heat, work, and efficiency, and analyzes the physical mechanism underlying the efficiency enhancement. Section~\ref{sec:numerics} presents the numerical tests of the mechanism. The appendices give the response formulas and the details needed to reproduce the calculations.

\section{Interaction model and quantum Otto cycle}\label{sec:model}

\subsection{Initial state and interaction model}

We consider a UQOE comprising two UDW detectors, labeled $A$ and $B$, that jointly constitute the working medium. Each detector is a two-level system with ground and excited states $|g_D\rangle$ and $|e_D\rangle$, respectively, where $D\in\{A,B\}$. We define the excited-state projector $n_D=|e_D\rangle\langle e_D|$ and the operator $S_D^z=2n_D-I_D$, where $I_D$ is the identity operator on detector $D$. The detectors have no direct coupling and share the same instantaneous energy gap $\omega$, which is modulated during the isolated strokes. Related single-detector UQOE models are discussed in Refs.~\cite{Mukherjee2022Unruh,Gallock2023Relativistic,Gallock-Yoshimura2023Quantum,misra2024black,Moustos2025Quantum,Sarkar2025Relativistic}.

The detectors are initially prepared in an uncorrelated product state, diagonal in the ordered basis $\{|gg\rangle,|eg\rangle,|ge\rangle, |ee\rangle\}$
\begin{equation}
	\rho_0=\diag(q_1,q_2,q_3,q_4),
	\label{eq:rho0}
\end{equation}
where $(q_1,q_2,q_3,q_4)=((1-p_A)(1-p_B),p_A(1-p_B),(1-p_A)p_B,p_Ap_B)$. We restrict the initial excited-state populations to $0<p_A<p_B<1/2$. Neither detector is then population-inverted, and the joint initial state is passive with respect to the free Hamiltonians considered below.

For detectors following different trajectories, their proper times must be related before the energy of the composite system can be expressed using a common time parameter. With respect to a common evolution parameter $t_*$, the free Hamiltonian is \cite{Dicke1954Coherence,Arias2015Boundary}
\begin{equation}
	H_{t_*}=\frac{\omega}{2}\left(\frac{\dd\tau_A}{\dd t_*}S_A^z\otimes\id_B+\frac{\dd\tau_B}{\dd t_*}\id_A\otimes S_B^z\right).
\end{equation}
Choosing detector $A$'s proper time as the reference yields
\begin{equation}
	H(\omega,b)=\frac{\omega}{2}(S_A^z\otimes\id_B+b\id_A\otimes S_B^z),\qquad
	b=\frac{\dd\tau_B}{\dd\tau_A}.
	\label{eq:H}
\end{equation}
In the specified product basis, the corresponding energy eigenvalues are $\frac{\omega}{2}\{-(1+b),\,1-b,\,b-1,\,1+b\}$. For $0<b<1$, the energy eigenstates appear in increasing order as $|gg\rangle$, $|ge\rangle$, $|eg\rangle$, and $|ee\rangle$. The population ordering $q_1>q_3>q_2>q_4$ therefore establishes passivity of the initial state. At $b=1$, the two single-excitation levels are degenerate, but the same product basis remains applicable. We retain the energy-zero convention of Eq.~\eqref{eq:H} throughout. The contributions associated with changing $b$ are distinguished from the work associated with modulating the gap $\omega$.

During each isochoric stroke, both detectors undergo uniform proper acceleration in the same direction, following trajectories in the right Rindler wedge:
\begin{equation}
	t_D=a_D^{-1}\sinh u_D,\quad x_D=a_D^{-1}\cosh u_D,\quad z_A=0,\quad z_B=L,
	\label{eq:worldlines}
\end{equation}
where $u_D=a_D\tau_D$ is the dimensionless boost time of detector $D$. On equal boost-time slices, $u_A=u_B=u$, implying $b=a_A/a_B$. This is the proper-time relation used in the supplementary material of Ref.~\cite{Kane2021Entangled}. The relation follows from the common boost-time foliation and should not be interpreted as a fixed proper-time ratio between events at equal Minkowski coordinate times. For the hot and cold isochoric strokes, we define
\begin{equation}
	\alpha_H=\frac{a_{A2}}{a_{B2}},\qquad
	\alpha_C=\frac{a_{A4}}{a_{B4}},\qquad
	0<\alpha_H,\alpha_C<1.
	\label{eq:alphas}
\end{equation}
Thus, detector $A$ has the smaller proper acceleration in both isochoric strokes. The subscripts $2$ and $4$ denote the hot (heat-absorption) and cold (heat-rejection) strokes, respectively. During the isolated adiabatic strokes, the detectors have a common inertial velocity and $b=1$. Equal acceleration is recovered only as a limiting case of the unequal-acceleration model.

Both detectors are linearly coupled to a massless scalar field prepared in the Minkowski vacuum of $(3+1)$-dimensional spacetime. In the interaction picture, using a common ordering time, the interaction Hamiltonian is
\begin{equation}
	\begin{aligned}\label{eq:interaction}
		H_I(u)&=\lambda\sum_{D=A,B}\frac{1}{a_D}\chi_D(u/a_D)\mu_D(u/a_D)\phi(x_D(u)),\\
		\mu_D(\tau_D)&=\sigma_D^+e^{\ii\omega\tau_D} +\sigma_D^-e^{-\ii\omega\tau_D}.
	\end{aligned}
\end{equation}
Here, $\lambda$ denotes the detector--field coupling strength, assumed sufficiently weak for a perturbative treatment. The relation $\tau_D=u/a_D$ accounts for the factor $1/a_D$ arising from reparametrization of the interaction. The operators $\sigma_D^+$ and $\sigma_D^-$ raise and lower the detector's internal state, respectively, while $\omega$ is the energy gap. The factors $e^{\pm\mathrm{i}\omega\tau_D}$ describe the free internal evolution, and $\phi(x_D(u))$ is the field operator evaluated on detector $D$'s trajectory. Eq.~\eqref{eq:interaction} therefore specifies a local detector--field coupling for each trajectory.

The switching function $\chi_D(\tau_D)$ controls the time dependence of this interaction. We adopt Gaussian switching functions \cite{brenna2016anti,Garay2016Thermalization,li2018would,wu2023conditions},
\begin{equation}
	\chi_D(\tau_D)=\exp\left[-\frac{(a_D\tau_D)^2}{2s_j^2}\right],
	\quad s_j = \Delta_D a_D,\quad j=H,C.
	\label{eq:switch}
\end{equation}
The dimensionless width $s_j$ is taken to be the same for both detectors during a given interaction stroke. $\Delta_D$ is the effective interaction time width between the detector and the field.

During each isochoric stroke, the detector pair exchanges energy with the field. In the adopted thermodynamic convention, the state change at fixed free Hamiltonian is accounted for as heat. The reduced-state evolution in the hot and cold strokes is represented perturbatively by
\begin{equation}
	\mathcal E_j(\rho)=\rho+\lambda^2R_j[\rho]+\ord(\lambda^4). \label{eq:map}
\end{equation}
The response matrices $R_j$ are defined without the common factor $\lambda^2$, and all control parameters are held fixed in the perturbative expansion. \ref{app:response} provides the complete matrix $R_j[\rho_0]$.

We assume that the field is initially in the Minkowski vacuum during each isochoric interaction. Thus the two contacts are modeled as independent vacuum environments. If the field is not reinitialized between strokes, correlations carried over from the first interaction generally produce additional contributions during the second. The model also does not specify a single finite trajectory continuously connecting the two accelerated interactions. Energetic costs associated with switching, field preparation, and motion control are not included in the reduced description.

\subsection{Four-stroke and state evolution}\label{sec:closure}

The two-detector working medium begins the cycle in the product state $\rho_0$ of Eq.~\eqref{eq:rho0}, with energy gap $\omega_1$ and relative proper-time weight $b=1$. We assume $0<\omega_1<\omega_2$. The cycle comprises two isolated gap-modulation strokes and two detector--field interaction strokes: expansion to the larger energy gap, the hot isochoric interaction, contraction to the smaller energy gap, and the cold isochoric interaction. During the isolated strokes, both detectors move inertially with the same velocity and are decoupled from the field. Thus $b=1$ and the energy-level populations remain fixed, although off-diagonal density-matrix elements may accumulate phases. The two field contacts use independent vacuum environments, as specified above, and the free evolution between them is treated explicitly. We adopt the convention that work is positive when performed \emph{on} the detector pair and heat is positive when absorbed by it.

\paragraph{Stroke I: adiabatic expansion.}
The detector--field coupling is switched off while both energy gaps increase from $\omega_1$ to $\omega_2$. The detectors move inertially at the common velocity $-v$, so $b=1$ and the free Hamiltonian is $H(\omega,1)$. Because this Hamiltonian is diagonal in a fixed basis, Hamiltonians at different times commute and the stroke induces no transitions between product energy eigenstates. Up to an irrelevant overall phase, the unitary is $U_{\rm I}=\exp[-\ii\vartheta_{\rm I}(n_A+n_B)]$, with $\vartheta_{\rm I}=\int_{\rm I}\omega(\tau_A)\,\dd\tau_A$. The diagonal initial state is therefore unchanged
\begin{equation}
	\rho_1=U_{\rm I}\rho_0U_{\rm I}^\dagger=\rho_0,
	\quad Q_{\rm I}=0.
	\label{eq:strokeIstate}
\end{equation}
The work performed on the pair through gap modulation is
\begin{equation}
	\begin{aligned}
		W_{\rm I}^{\rm on}&=\Tr\!\left[\rho_0\bigl(H(\omega_2,1)-H(\omega_1,1)\bigr)\right]\\
		&=(\omega_2-\omega_1)(p_A+p_B-1).
	\end{aligned}
	\label{eq:strokeIwork}
\end{equation}
The sign of an individual work term reflects the energy-zero convention in Eq.~\eqref{eq:H}.

\paragraph{Stroke II: hot isochoric interaction.}
The gap is held fixed at $\omega_2$, and the detectors interact with the field along uniformly accelerated trajectories with accelerations $a_{A2}$ and $a_{B2}$. The relative proper-time weight is $b=\alpha_H=a_{A2}/a_{B2}$. Along these trajectories, the velocity ranges from \(-v\) to \(v\) over a finite kinematic segment; the associated proper-time half-widths may be parametrized as \(\Delta_A = \frac{\operatorname{arctanh}(v)}{a_A}\) and \(\Delta_B = \frac{\operatorname{arctanh}(v)}{a_B}\), respectively. For the Gaussian interactions of Eq.~\eqref{eq:switch}, these quantities specify characteristic widths rather than sharp switching endpoints. To second order, the state after this interaction is
\begin{equation}
	\rho_2\equiv\rho_H=\mathcal E_H(\rho_1)
	=\rho_0+\lambda^2R_H[\rho_0]+\ord(\lambda^4).
	\label{eq:strokeIIstate}
\end{equation}
In the ordered basis of Eq.~\eqref{eq:rho0}, the full reduced density matrix reads
\begin{equation}
	\begin{aligned}
		\rho_H=&
		\begin{pmatrix}
			q_1+\lambda^2S_{1,H}&0&0&\lambda^2Y_{14,H}\\
			0&q_2+\lambda^2S_{2,H}&\lambda^2Y_{23,H}&0\\
			0&\lambda^2Y_{23,H}^*&q_3+\lambda^2S_{3,H}&0\\
			\lambda^2Y_{14,H}^*&0&0&q_4+\lambda^2S_{4,H}
		\end{pmatrix}\\&+\ord(\lambda^4).
		\label{eq:hotstatematrix}
	\end{aligned}
\end{equation}
The coefficients are given in~\ref{app:response} and obey $\sum_{k=1}^4S_{k,H}=0$. The diagonal terms describe changes in the level populations, whereas the off-diagonal terms describe complex field-induced coherences. Nonzero coherences alone do not establish entanglement, which must be evaluated using an appropriate state criterion. Since the free Hamiltonian in Eq.~\eqref{eq:H} is constant during this stroke, the absorbed heat is
\begin{equation}
	\begin{aligned}
		Q_H&=\Tr\!\left[H(\omega_2,\alpha_H)(\rho_H-\rho_1)\right]\\
		&=\lambda^2\Tr\!\left[H(\omega_2,\alpha_H)R_H[\rho_0]\right]
		+\ord(\lambda^4).
	\end{aligned}\label{eq:strokeIIheat}
\end{equation}
No work is performed during this stroke. The designation ``hot'' is operational: an admissible engine point must satisfy $Q_H>0$. For a uniformly accelerated detector, the transition probability satisfy the Unruh detailed-balance relation in the long-interaction-time limit. At finite duration, however, the response depends on the switching protocol and does not generally represent reaching a thermal equilibrium state \cite{brenna2016anti,Garay2016Thermalization,wu2023conditions}.

\paragraph{Stroke III: adiabatic compression.}
The detector--field coupling is switched off, and the detectors move inertially at their common velocity $v$, giving $b=1$. Both energy gaps decrease from $\omega_2$ to $\omega_1$. We denote the common free-evolution phase between the field contacts by $\vartheta$; for this stroke alone, $\vartheta=\int_{\rm III}\omega(\tau_A)\,\dd\tau_A$. The unitary evolution and detector state are
\begin{equation}
	\begin{aligned}	\label{eq:strokeIIIstate}
		U_\vartheta&=e^{-\ii\vartheta(n_A+n_B)}
		=\diag(1,e^{-\ii\vartheta},e^{-\ii\vartheta},e^{-2\ii\vartheta}),\\
		\rho_3&=U_\vartheta\rho_HU_\vartheta^\dagger
		=\rho_0+\lambda^2U_\vartheta R_H[\rho_0]U_\vartheta^\dagger+\ord(\lambda^4).
	\end{aligned}
\end{equation}
The populations remain unchanged, but the coherences may acquire dynamical phases: $Y_{14,H}$ transforms to $e^{2\ii\vartheta}Y_{14,H}$, whereas $Y_{23,H}$ is unchanged because $|eg\rangle$ and $|ge\rangle$ are degenerate during the stroke. The stroke exchanges no heat, $Q_{\rm III}=0$, and the work is
\begin{equation}
	W_{\rm III}^{\rm on}
	=\Tr\!\left[\rho_H\bigl(H(\omega_1,1)-H(\omega_2,1)\bigr)\right].
	\label{eq:strokeIIIwork}
\end{equation}
Using $\rho_H$ instead of $\rho_3$ in this trace is justified by the equality of their diagonal populations, not by an assumption that the coherences remain fixed.

\paragraph{Stroke IV: cold isochoric interaction.}
At the fixed gap $\omega_1$, the detectors couple to an independently prepared vacuum field, with accelerations $a_{A4}$ and $a_{B4}$ and relative proper-time weight $b=\alpha_C=a_{A4}/a_{B4}$. For the reverse kinematic process, in which the velocities range from \(v\) to \(-v\), the effective interaction times Eq.~\eqref{eq:switch} are given by \(\Delta_A = \frac{\operatorname{arctanh}(v)}{a_A}\) and \(\Delta_B = \frac{\operatorname{arctanh}(v)}{a_B}\), respectively. The input state is $\rho_3$, rather than $\rho_0$. Nevertheless, $\rho_3-\rho_0=\ord(\lambda^2)$, and the response is linear in the input state. Replacing $\rho_3$ by $\rho_0$ inside the term $\lambda^2R_C$ therefore changes the result only at order $\lambda^4$
\begin{equation}
	\begin{aligned}
		\rho_f&=\mathcal E_C(\rho_3)
		=\rho_3+\lambda^2R_C[\rho_3]+\ord(\lambda^4)\nonumber\\
		&=\rho_0+\lambda^2\left(
		U_\vartheta R_H[\rho_0]U_\vartheta^\dagger+R_C[\rho_0]\right)
		+\ord(\lambda^4).
		\label{eq:strokeIVstate}
	\end{aligned}
\end{equation}
The heat exchanged during the cold stroke is
\begin{equation}
	\begin{aligned}
		Q_C&=\Tr\!\left[H(\omega_1,\alpha_C)(\rho_f-\rho_3)\right]\\
		&=\lambda^2\Tr\!\left[H(\omega_1,\alpha_C)R_C[\rho_0]\right]
		+\ord(\lambda^4).
	\end{aligned}	\label{eq:strokeIVheat}
\end{equation}
Heat rejection requires $Q_C<0$, which must be established from the response rather than assumed from the smaller gap or acceleration. The cold-stroke parameters must also ensure return to the initial state in Eq.~\eqref{eq:rho0}.

\begin{figure*}[t]
	\centering
	\includegraphics[width=0.75\textwidth]{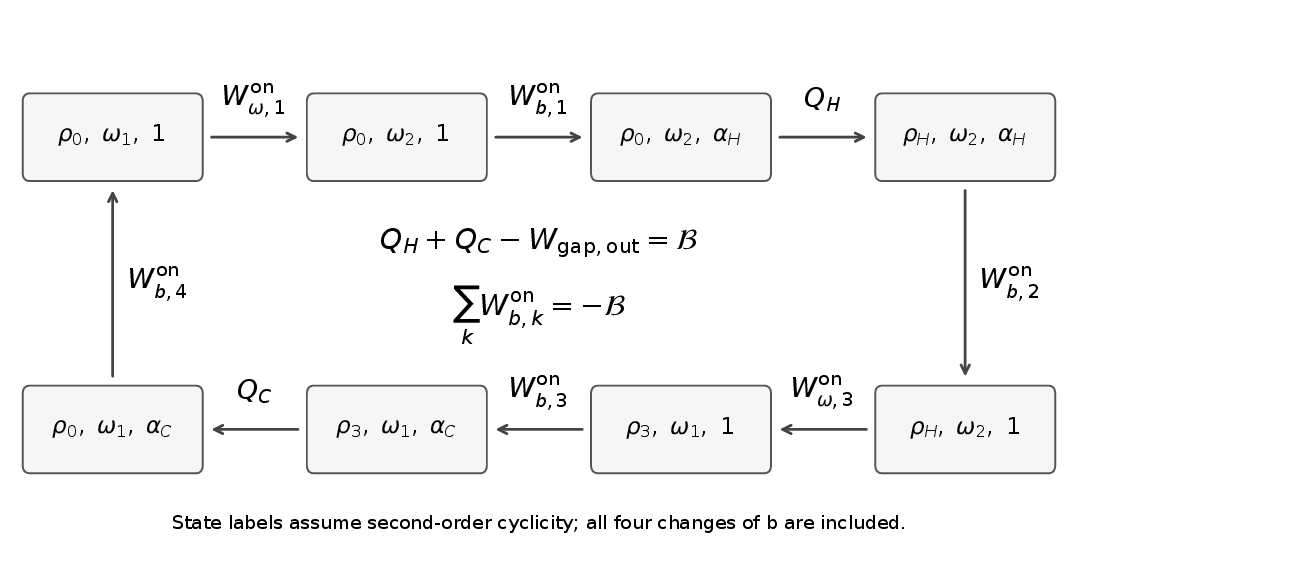}
	\caption{Parameter changes during the heat-engine cycle. Each box lists $(\rho,\omega,b)$. The state $\rho_3=U_\vartheta\rho_HU_\vartheta^\dagger$ has the local response terms and the coherence terms. Returning to $\rho_0$ assumes that Eq.~\eqref{eq:closure} is satisfied.}
	\label{fig:ledger}
\end{figure*}

The full protocol includes four changes of the relative proper-time weight: $1\to\alpha_H$ before the hot stroke, $\alpha_H\to1$ after it, $1\to\alpha_C$ before the cold stroke, and $\alpha_C\to1$ afterward. These changes are idealized boundaries in the reduced model and are assumed not to modify the detector state. All four are displayed in figure~\ref{fig:ledger}. 

The detector-state evolution over the four strokes is summarized by
\begin{equation}
	\rho_0\xrightarrow{\mathrm{I}}\rho_1=\rho_0
	\xrightarrow{\mathrm{II}}\rho_H
	\xrightarrow{\mathrm{III}}U_\vartheta\rho_HU_\vartheta^\dagger=\rho_3
	\xrightarrow{\mathrm{IV}}\rho_f.
	\label{eq:cyclestates}
\end{equation}
State closure through the retained perturbative order requires
\begin{equation}
	R_f\equiv U_\vartheta R_H[\rho_0]U_\vartheta^\dagger+R_C[\rho_0]=0,
	\quad \rho_f=\rho_0+\ord(\lambda^4).
	\label{eq:closure}
\end{equation}
To display this matrix condition explicitly, we define $d_k=S_{k,H}+S_{k,C}$ for $k=1,\ldots,4$, together with $z_{14}=e^{2\ii\vartheta}Y_{14,H}+Y_{14,C}$ and $z_{23}=Y_{23,H}+Y_{23,C}$. The closure residual is then
\begin{equation}
	R_f=
	\begin{pmatrix}
		d_1&0&0&z_{14}\\
		0&d_2&z_{23}&0\\
		0&z_{23}^*&d_3&0\\
		z_{14}^*&0&0&d_4
	\end{pmatrix}=0.
	\label{eq:cyclematrix}
\end{equation}
Therefore, the diagonal population changes must cancel, $d_k=0$, with $\sum_k d_k=0$ following from trace preservation. Both complex coherence terms must also cancel
\begin{equation}
	e^{2\ii\vartheta}Y_{14,H}+Y_{14,C}=0,\quad
	Y_{23,H}+Y_{23,C}=0.
	\label{eq:cohclosure}
\end{equation}
Both the real and imaginary parts of these equations must vanish. Cancellation of the diagonal populations alone, or of only the real parts of the coherences, is insufficient for state closure. Conversely, return to a separable initial state does not exclude the generation of entanglement at an intermediate stage.

A closed state cycle is classified as an engine only after the heat and work satisfy the appropriate sign conditions, including $Q_H>0$, $Q_C<0$, and positive work output. The output associated with energy gap modulation is $W_{\rm gap,out}=-(W_{\rm I}^{\rm on}+W_{\rm III}^{\rm on})$. Energy balance and the distinction between work and additional control contributions are examined in the next section.

\section{Heat, work, and the origin of efficiency enhancement}\label{sec:energy}

For the two-detector working medium, the internal energy associated with the free Hamiltonian $H$ is $E=\Tr(\rho H)$. Its differential separates into state and Hamiltonian contributions \cite{Arias2017Unruh,Gray2018Scalar}
\begin{equation}
	\begin{aligned}
		\dd E&=\Tr(H\,\dd\rho)+\Tr(\rho\,\dd H),\\
		\delta Q&=\Tr(H\,\dd\rho),\quad
		\delta W^{\rm on}=\Tr(\rho\,\dd H).
	\end{aligned}\label{eq:firstlaw}
\end{equation}
Here, $\Tr$ is the trace over the two-detector Hilbert space, $\dd\rho$ denotes a change in the state, and $\dd H$ denotes a change in the free Hamiltonian. We identify $\delta Q=\Tr(H\,\dd\rho)$ with heat absorbed through changes in the state at fixed Hamiltonian and $\delta W^{\rm on}=\Tr(\rho\,\dd H)$ with work performed on the medium through controlled Hamiltonian changes. This is the heat--work convention employed in Refs.~\cite{Kane2021Entangled,Barman2021Constructing}.

During an isochoric stroke, both $\omega$ and $b$ are held fixed, so $\dd H=0$ and the change in the detector energy is assigned to heat
\begin{equation}
	\dd E=\delta Q=\Tr(H\,\dd\rho).
\end{equation}
In the present model, this change arises from the detector--field interaction. During an isolated adiabatic stroke, by contrast, the detectors are decoupled from the field and the energy gaps are varied externally. The energy level populations remain constant, and the corresponding energy change is identified as work
\begin{equation}
	\dd E=\delta W^{\rm on}=\Tr(\rho\,\dd H).
\end{equation}
This decomposition accounts for changes in the expectation value of the detectors' free Hamiltonian. It does not include the additional energetic costs of switching the detector--field interaction, resetting the field environment, or controlling the detector trajectories.

Unless otherwise indicated, the following heat and work expressions retain terms through second order in $\lambda$. For a state closed cycle, Eqs.~\eqref{eq:H}, \eqref{eq:strokeIwork}, \eqref{eq:strokeIIheat}, \eqref{eq:strokeIIIwork}, and \eqref{eq:strokeIVheat} give
\begin{align}
	Q_H&=\omega_2(X_A+\alpha_HX_B),\quad
	Q_C=-\omega_1(X_A+\alpha_CX_B),\nonumber\\
	W_{\rm gap,out}&=(\omega_2-\omega_1)(X_A+X_B).
	\label{eq:W}
\end{align}
where $X_A=\lambda^2\Tr((n_A\otimes\id_B) R_H[\rho_0])$ and $X_B=\lambda^2\Tr((\id_A\otimes n_B) R_H[\rho_0])$ denote the hot-stroke changes in the local excited-state populations, with $\Tr(\mathbb{I} R_H[\rho_0])=0$. In the basis $\{|gg\rangle, |eg\rangle, |ge\rangle, |ee\rangle\}$, the local projectors are $n_A\otimes\id_B=\mathrm{diag}(0, 1, 0, 1)$ and $\id_A\otimes n_B=\mathrm{diag}(0, 0, 1, 1)$. In particular, detector $B$ contributes $\alpha_H\omega_2X_B$ to the heat absorbed during the hot stroke, whereas its contribution to the output from the two gap-modulation strokes is $(\omega_2-\omega_1)X_B$. This difference follows because the hot stroke has relative time weight $\alpha_H$, while the isolated gap strokes have $b=1$. Positive $W_{\rm gap,out}$ denotes extracted work. For $Q_H>0$, we define $r=\omega_1/\omega_2$ and $\eta_0=1-r$, giving
\begin{equation}
	\eta_{\rm gap}=\frac{W_{\rm gap,out}}{Q_H} =\eta_0\frac{X_A+X_B}{X_A+\alpha_HX_B},
	\label{eq:eta}
\end{equation}
The subscript ``gap'' emphasizes that the numerator includes work from gap modulation only. Eq.~\eqref{eq:eta} is therefore a ratio of specified energy exchanges; it represents an admissible cycle efficiency within the reduced energy accounting only when state closure, energy balance, and the required heat--work signs have been verified.

\subsection{Energy balance under changes of the relative proper-time weight}

The condition Eq.~\eqref{eq:closure} that only the UQOE state returns to its initial state does not ensure that the energy satisfies Eq.~\eqref{eq:W}. The difference between the heat and the work output is
\begin{equation}
	\mathcal{B} \equiv Q_H+Q_C-W_{\rm gap,out}
	=\bigl[\omega_2(\alpha_H-1)-\omega_1(\alpha_C-1)\bigr]X_B.
	\label{eq:B}
\end{equation}
If changes in $b$ are implemented as Hamiltonian controls, each boundary contributes $W_{b,k}^{\rm on}=\omega(b_{\rm after}-b_{\rm before})\langle S_B^z\rangle/2$. The four boundaries are shown in figure.~\ref{fig:ledger}. Their combined contribution is
\begin{equation}
	\sum_{k=1}^4W_{b,k}^{\rm on}
	=\bigl[\omega_2(1-\alpha_H)+\omega_1(\alpha_C-1)\bigr]X_B=-\mathcal{B}.
	\label{eq:boundary}
\end{equation}
Thus the output including these control contributions is $W_{\rm gap,out}+\mathcal{B}=Q_H+Q_C$. Alternatively, if $b$ encodes a change in the reference time used to define energy, these terms account for differences in time-weighted energies and should not automatically be identified with the mechanical cost of accelerating the detectors.

For $X_B\ne0$, requiring $\mathcal{B}=0$ imposes
\begin{equation}
	\alpha_C=1-\frac{1-\alpha_H}{r},\quad
	1-r<\alpha_H<1,\quad 0<\alpha_C<\alpha_H.
	\label{eq:balance}
\end{equation}
The inequalities follow from $0<r<1$ and $0<\alpha_C<1$. This energy-balance constraint is related to those discussed in earlier UQOE studies \cite{Arias2017Unruh,Gray2018Scalar,Kane2021Entangled,Barman2021Constructing}; it is distinct from the state closure condition emphasized, in particular, in Ref.~\cite{Barman2021Constructing}. When state closure and $\mathcal{B}=0$ both hold, $W_{\rm gap,out}=Q_H+Q_C$. If, in addition, $Q_H>0$, $Q_C<0$, and $W_{\rm gap,out}>0$, then $\eta_{\rm gap}$ is the efficiency associated with the specified energy accounting. If the same nonunit acceleration ratio is imposed in both isochoric strokes, $\alpha_H=\alpha_C =\alpha\ne1$, the condition $\mathcal{B}=0$ instead requires $X_B=0$; Eq.~\eqref{eq:eta} then reduces to $\eta_0$ whenever the ratio is defined. Thus imposing a common nonunit acceleration ratio introduces an additional restriction with a direct energetic consequence.

\subsection{Local population response and relative proper-time weight}

Subtracting $\eta_0$ from Eq.~\eqref{eq:eta} yields
\begin{equation}
	\Delta\eta_{\rm gap}=\eta_{\rm gap}-\eta_0
	=\eta_0\frac{(1-\alpha_H)X_B}{X_A+\alpha_HX_B}.
	\label{eq:enhance}
\end{equation}
For $Q_H>0$, $0<\alpha_H<1$, and $0<r<1$, we obtain
\begin{equation}
	\eta_{\rm gap}>\eta_0\quad\Longleftrightarrow\quad X_B>0.
	\label{eq:criterion}
\end{equation}
The physical origin of this criterion is explicit in Eq.~\eqref{eq:eta}. The factor $\alpha_H$ arises from expressing both detectors' energy changes using detector $A$'s proper time, whereas $X_B$ is the population change of detector $B$ induced during the hot stroke. For $\alpha_H<1$ and $X_B>0$, this population change contributes less to absorbed heat than to work because it carries different weights in the two quantities. The efficiency therefore exceeds $\eta_0$. Both unequal proper-time weighting and a positive local response are required. The enhancement vanishes if $\alpha_H=1$ or if $X_B=0$ with nonzero absorbed heat; for $X_B<0$, the same weighting reduces the efficiency ratio.

For $S=X_A+X_B>0$, define $f=X_B/S$ and $\zeta=(1-\alpha_H)f$. Combining Eqs.~\eqref{eq:eta} and \eqref{eq:balance}, we obtain
\begin{equation}
	\eta_{\rm gap}=\frac{1-r}{1-\zeta},\quad Q_H=\omega_2S(1-\zeta),\quad Q_C=-\omega_2S(r-\zeta).
\end{equation}
An enhanced operating point satisfying energy balance and the heat-sign conditions must therefore obey $0<\zeta<r$.

\subsection{Entanglement does not contribute directly}
\label{sec:dephasing}

The independence of energy exchange from coherence under the adopted diagonal Hamiltonian can be established without introducing an entanglement measure. Let $\mathcal D$ denote complete dephasing
\begin{equation}
	\mathcal D(\rho)=\sum_{n\in\{gg,eg,ge,ee\}} |n\rangle\langle n|\rho|n\rangle\langle n|.
	\label{eq:dephasing}
\end{equation}
At fixed populations and control parameters, removing the off-diagonal matrix elements leaves the internal energy, heat, and work expressions unchanged. Indeed, $H=\mathcal D(H)$ implies
\begin{equation}
	\Tr\!\left[H\bigl(\rho-\mathcal D(\rho)\bigr)\right]=0.
	\label{eq:diagonalenergy}
\end{equation}
To make this distinction explicit, consider a fixed initial state $\rho_0$ and define the comparison quantities
\begin{equation}
	\begin{aligned}
		Q_H[\rho_2;\rho_0]=&\Tr[H(\omega_2,\alpha_H)(\rho_2-\rho_0)],\\
		W_{\rm gap,out}[\rho_2;\rho_0]
		=&(\omega_2-\omega_1)\\&\times\Tr[(n_A\otimes\id_B+\id_A\otimes n_B)(\rho_2-\rho_0)].
		\label{eq:comparison}
	\end{aligned}
\end{equation}
The second line evaluates work for a chosen comparison state $\rho_2$ following the hot stroke. Since $\rho_0$ is diagonal, Eq.~\eqref{eq:diagonalenergy} gives
\begin{equation}
	\begin{aligned}
		Q_H[\rho_2;\rho_0]&=Q_H[\mathcal D(\rho_2);\rho_0],\\
		W_{\rm gap,out}[\rho_2;\rho_0]&=W_{\rm gap,out}[\mathcal D(\rho_2);\rho_0].
	\end{aligned}   \label{eq:dephasedenergies}
\end{equation}
The ratio of these quantities is likewise invariant under dephasing whenever the absorbed heat is nonzero. This is an exact algebraic consequence of the diagonal energy observables. In particular, for $\rho_2=\rho_0+\lambda^2R_H[\rho_0]$, it yields the second-order relation used here. The post-hot-stroke coherence matrix elements are $\lambda^2Y_{14,H}$ and $\lambda^2Y_{23,H}$, not the response coefficients $Y_{14,H}$ and $Y_{23,H}$ alone.

The diagonal state obtained by dephasing is separable, whereas the state with off-diagonal terms can be entangled. Thus coherence need not contribute directly to the chosen energy observables even though the state closure condition constrains it. State closure and energy balance are distinct requirements; satisfying either one alone does not guarantee the other. Moreover, the local-response simplification relies on the chosen initial preparation. For the product state considered here, the population changes are given by Eq.~\eqref{eq:localpop}, whereas initially coherent or entangled states can exhibit different population dynamics, including contributions involving cross-detector correlations \cite{Kane2021Entangled,Barman2021Constructing}.

\subsection{Detector separation, local excitation change}
\label{sec:separation}

For the initial product state of Eq.~\eqref{eq:rho0}, the local excitation change is
\begin{equation}
	X_D=\lambda^2 [(1-p_D)U_{D,H}-p_DV_{D,H}],
	\label{eq:localpop}
\end{equation}
where $U_{D,H}$ and $V_{D,H}$ denote the excitation and deexcitation response coefficients, respectively. These depend on the pullback of the field two-point function to detector $D$'s own trajectory. At fixed local controls, changing the transverse detector separation modifies the nonlocal coherence terms but does not alter either detector's local response. Hence, with accelerations, energy gaps, switching functions, phases, and initial populations held fixed,
\begin{equation}
	\partial_LX_A=\partial_LX_B=0.
	\label{eq:localL}
\end{equation}
This property follows from the locality of the response coefficients for the product initial state in Eq.~\eqref{eq:rho0}.

Writing $Q_H=\lambda^2q_H^{(2)}+\ord(\lambda^4)$, and denoting the leading work coefficient and efficiency ratio by $w^{(2)}$ and $\eta^{(0)}$, respectively, the population-dependent expressions derived from Eq.~\eqref{eq:W} satisfy
\begin{equation}
	\partial_L q_H^{(2)}=\partial_L w^{(2)} =\partial_L\eta^{(0)}=0,\quad
	\eta^{(0)}=\eta_0\frac{X_A+X_B}{X_A+\alpha_HX_B}.
	\label{eq:Ltheorem}
\end{equation}
By contrast, the off-diagonal coherences $Y_{14}(L)$ and $Y_{23}(L)$ generally depend on the transverse separation $L$. Within the second-order approximation, heat, gap work, and their ratio are independent of $L$ only when the local controls and initial populations are fixed, and state closure is not imposed.

However, to constitute a complete thermodynamic cycle, we must consider the state closure condition, because the nonlocal coherences enter the conditions in Eq.~\eqref{eq:cohclosure}. To restore state closure, other control parameters may need to vary with separation. Consequently, the achievable efficiency can acquire an indirect dependence on $L$. If the parameters are collected in $\boldsymbol\theta(L)$, the derivative along the closed-cycle branch is
\begin{equation}
	\frac{\dd\eta_{\rm gap}^{(0)}}{\dd L} =\sum_k\frac{\partial\eta_{\rm gap}^{(0)}}{\partial\theta_k} \frac{\dd\theta_k}{\dd L}, \label{eq:totalderivative}
\end{equation}
which need not vanish.

\subsection{Entanglement generation}
\label{sec:limits}

For a physical two-qubit $X$ state with diagonal elements $e_i$ and off-diagonal elements $f_{14},f_{23}$, the concurrence is \cite{Wootters1998Entanglement,yu2005evolution,yu2006sudden}
\begin{equation}
	C=2\max\{0,|f_{14}|-\sqrt{e_2e_3},|f_{23}|-\sqrt{e_1e_4}\}.
	\label{eq:C}
\end{equation}
This expression applies to a positive semidefinite density matrix; a truncated matrix that is not positive cannot be treated as an exact physical state. 

For the initial state in Eq.~\eqref{eq:rho0}, define
\begin{equation}
	m_0=\lambda_{\min}(\rho_0^{T_B}) =\min(p_A,1-p_A)\min(p_B,1-p_B)>0.
\end{equation}
Assume that the exact reduced state is positive and admits a controlled expansion of the form
\begin{equation}
	\begin{aligned}
		\rho_H&=\rho_0+\lambda^2R_H[\rho_0]+E_4,\quad \|E_4^{T_B}\|_{\rm op}\leq\epsilon_4,\\ K&=\|(R_H[\rho_0])^{T_B}\|_{\rm op}.
	\end{aligned}\label{eq:remainder}
\end{equation}
For any unit vector, the expectation value of the last two terms is bounded below by $-\lambda^2K-\epsilon_4$. Minimization over unit vectors therefore yields
\begin{equation}
	\lambda_{\min}(\rho_H^{T_B})\geq m_0-\lambda^2K-\epsilon_4.
\end{equation}
For two-qubit states, positivity under partial transposition (PPT) is necessary and sufficient for separability \cite{Peres1996Separability,Horodecki1996Separability}. A sufficient condition is therefore
\begin{equation}
	\lambda^2K+\epsilon_4<m_0\quad\Longrightarrow\quad\rho_H
	\text{ is separable}.
	\label{eq:PPTbound}
\end{equation}
If $\epsilon_4\leq M\lambda^4$ with a known constant $M>0$, this condition is guaranteed when
\begin{equation}
	\lambda^2<\frac{\sqrt{K^2+4Mm_0}-K}{2M}.
\end{equation}
For $M=0$, the condition becomes $\lambda^2<m_0/K$ when $K>0$. These are sufficient, not necessary, criteria.

For the $X$-structured matrix $R_H[\rho_0]$ with diagonal elements $S_i$, the eigenvalues of its partial transpose are
\begin{equation}
	\begin{aligned}
		&\frac{S_1+S_4\pm\sqrt{(S_1-S_4)^2+4|Y_{23}|^2}}{2},\\
		&\frac{S_2+S_3\pm\sqrt{(S_2-S_3)^2+4|Y_{14}|^2}}{2}.
	\end{aligned}\label{eq:Keigen}
\end{equation}
The largest absolute eigenvalue determines $K$.

The condition $\lambda^2K+\epsilon_4<m_0$ guarantees separability of the post-hot-stroke state. For a fixed
full-rank product initial state, this condition can be satisfied at sufficiently weak coupling if the higher-order remainder is appropriately bounded. Notably, the independence of heat and work from coherence follows directly from the diagonal free Hamiltonian, regardless of the separability of the state.

\section{Numerical results}\label{sec:numerics}

We now use numerical calculations to examine the mechanism behind work efficiency enhancement and its relation to coherence. First, we construct closed cycles and compare their efficiency with the conventional Otto benchmark. Simultaneously, we analyze the absorbed heat and the entanglement properties of the post-hot-stroke state. Second, we investigate the impact of the detector separation on the engine's efficiency. Throughout, the state used for the entanglement test is the hot-stroke output $\rho_H$, rather than the final state after heat rejection. Numerical methods and independent consistency checks are described in~\ref{app:numericalprocedure}.

\begin{figure}[t]
	\centering
	\includegraphics[width=0.45\textwidth]{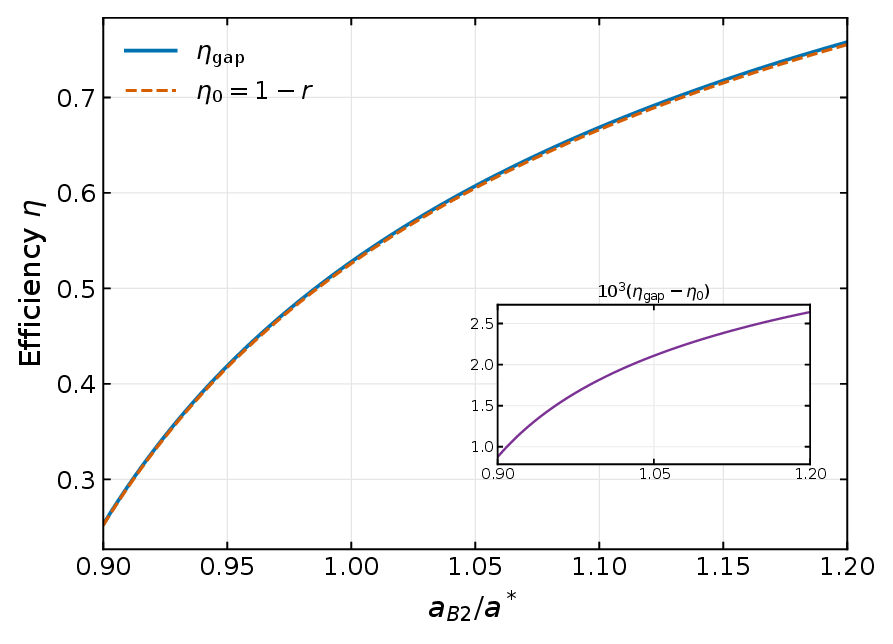}
	\caption{One hundred closed UQOE cycles with enhanced efficiency. The fixed parameters are
		given in Eq.~\eqref{eq:fixedcontrols}; $\lambda^2=10^{-3}$. The blue solid curve is the UQOE efficiency $\eta_{\rm gap}=W_{\rm gap,out}/Q_H$. The orange dashed curve is $\eta_0=1-\omega_1/\omega_2$, evaluated at the same gaps for each point.
		Inset: their positive difference multiplied by $10^3$. 
		Numerical checks are listed in Table~\ref{tab:verification}.}
	\label{fig:efficiency_comparison}
\end{figure}

\subsection{Closed-cycle operating points: enhanced efficiency without entanglement}
\label{sec:branch}

We consider the two-detector working medium over all four strokes and sample 100 equally spaced values of the hot-stroke acceleration in $0.9\leq a_{B2}/a^*\leq1.2$, where $a^*=1$ defines a reference acceleration. The hot acceleration ratio, Gaussian widths, and transverse separation are fixed at $\alpha_H\simeq0.99310$, $s_H=s_C\simeq0.49928$, and $a^*L\simeq1.96027$. For each $a_{B2}$, we determine the energy gaps and cold-stroke acceleration parameters by imposing state closure and energy balance. The initial populations and free-evolution phase are determined consistently with these conditions. Each point is thus an independently checked closed cycle, not an acceleration scan at fixed initial preparation.

For each accepted cycle, we calculate $Q_H$, $Q_C$, and $W_{\rm gap,out}$ from the local population changes and evaluate $\eta_{\rm gap}$. We independently test the second-order post-hot-stroke matrix $\rho_H^{[2]}$ for positivity and separability at $\lambda^2=10^{-3}$. All 100 sampled cycles absorb heat, reject heat, and produce positive gap work, while satisfying state closure and $\mathcal{B}=0$.

Figure~\ref{fig:efficiency_comparison} shows that the efficiency exceeds the conventional Otto value throughout the sampled range, although the enhancement is quantitatively small. The two curves nearly overlap; the inset displays their difference multiplied by $10^3$. Before rescaling, the positive difference ranges from $8.76\times10^{-4}$ to $2.64\times10^{-3}$. 

\begin{figure*}[t]
	\centering
	\includegraphics[width=0.9\textwidth]{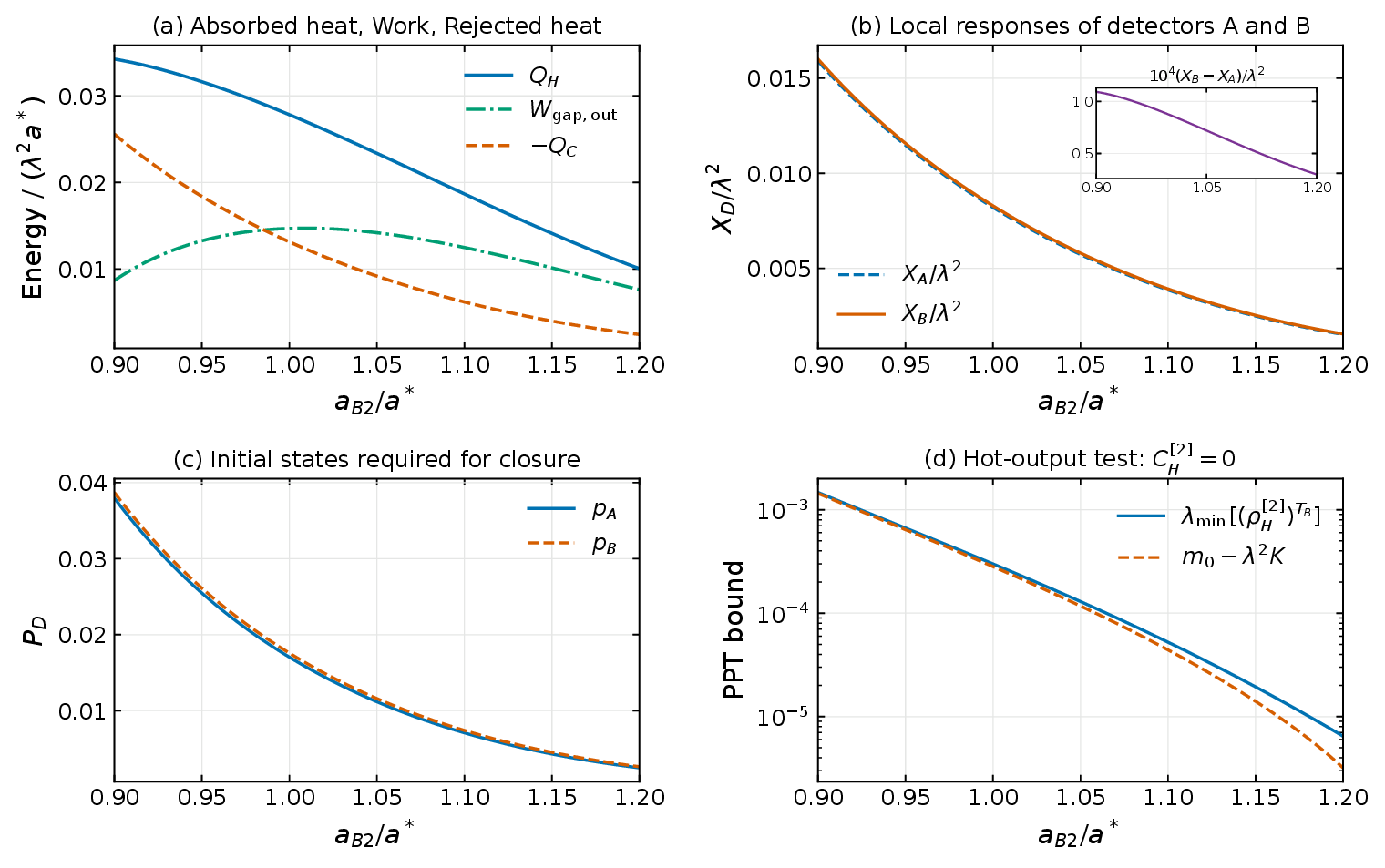}
	\caption{Variations of the relevant physical quantities during the heat absorption stage with respect to the acceleration. The values of the other fixed parameters are the same as those in Fig. \ref{fig:efficiency_comparison}, and numerical checks are listed in Table~\ref{tab:verification}.
		All horizontal axes, including insets, span $0.9\leq a_{B2}/a^*\leq1.2$.
		(a) Blue denotes absorbed heat $Q_H$, green dash-dotted denotes output gap work $W_{\rm gap,out}$, and orange dashed denotes rejected heat $-Q_C>0$, in units of $\lambda^2a^*$.
		(b) The blue dashed and orange solid curves are $X_A/\lambda^2$ and $X_B/\lambda^2$, respectively, where $X_D$ is the increase in UQOE $D$'s excited-state probability.  The inset resolves their small difference as $10^4(X_B-X_A)/\lambda^2$.
		(c) The initial excited-state probabilities $p_A$ (blue solid) and $p_B$ (orange dashed).
		(d) Blue solid: minimum eigenvalue of $(\rho_H^{[2]})^{T_B}$; orange dashed: $m_0-\lambda^2K$ (logarithmic scale). Colors identify quantities, not entanglement; $C_H^{[2]}=0$ at every
		point. }
	\label{fig:branch}
\end{figure*}

In figure~\ref{fig:branch}, all panels are plotted against the hot-stroke acceleration $a_{B2}/a^*$ of detector $B$. Panel (a) resolves the energy exchanges entering the work efficiency, each normalized by $\lambda^2a^*$. The blue curve equals the sum of the other two because $\mathcal{B}=0$. Along the closed-cycle branch, the absorbed and rejected heats decrease, while the output gap work first increases and then decreases. An increase in efficiency therefore does not necessarily imply greater work output per cycle, since the absorbed heat also changes.

Panel (b) shows the local excited-state population changes. Both responses are positive, with detector $B$ exhibiting a slightly larger response along this branch. Crucially, the enhancement criterion is $X_B>0$, not $X_B>X_A$: the same $X_B$ contributes to absorbed heat with weight $\alpha_H<1$ and to work with unit weight. This unequal weighting explains the positive efficiency enhancement even as both local responses decrease.

Panel (c) displays the initial excited-state populations $p_A$ and $p_B$, rather than their hot-stroke changes. They satisfy $0<p_A<p_B<1/2$ and decrease along the branch. The initial states are passive under the adopted Hamiltonians but are not identical across operating points; the other solved controls also vary with $a_{B2}$.

Panel (d) examines the post-hot-stroke state's separability. The blue curve gives the minimum eigenvalue of $(\rho_H^{[2]})^{T_B}$. Its positivity, together with positivity of $\rho_H^{[2]}$, establishes separability of all 100 second-order truncated hot-output states, with $C_H^{[2]}=0$. The orange curve shows $m_0-\lambda^2K$, the allowed norm of an uncomputed higher-order remainder under the sufficient condition in Eq.~\eqref{eq:PPTbound}. 

These results show that enhanced efficiency can coexist with separability of the second-order post-hot-stroke state. The enhancement is determined by the local population responses and their relative proper-time weighting, rather than by a direct energetic contribution from generated entanglement.

\begin{figure*}[t]
	\centering
	\includegraphics[width=0.9\textwidth]{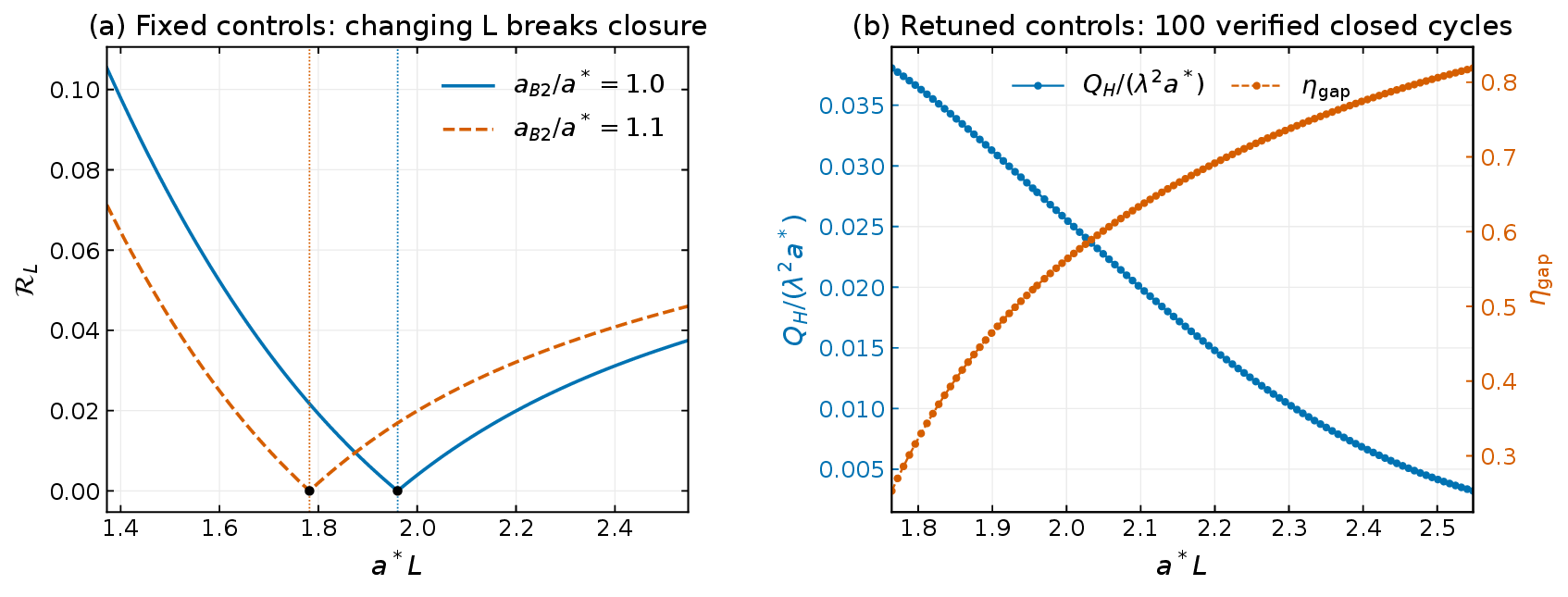}
	\caption{Two different tests of the transverse distance $L$ between two detector trajectories, perpendicular to their acceleration direction. The horizontal axes show $a^*L$. The two curves in (a) correspond to different values of $a_{B2}$, and the black markers denote the parameter points that satisfy the cycle conditions; (b) Heat $Q_H/(\lambda^2a^{*})$ (blue, left axis) and efficiency $\eta_{\rm gap}$ (orange, right axis) along the branch obtained by retuning the cycle parameters and initial populations at each separation. Each marker denotes an independently verified cycle satisfying state closure, energy balance, the required heat and work signs, and the state-positivity checks.}
	\label{fig:separation}
\end{figure*}

\subsection{Changing detector separation}
\label{sec:fixedseparation}

We next study the effect of transverse detector separation, starting from the reference closed-cycle operating point specified in Eq.~\eqref{eq:reference}. In the fixed-control test, the initial populations, energy gaps, both sets of accelerations, Gaussian switching widths, and free-evolution phase are held fixed while only the separation $L$ is varied. We recalculate the response matrices after the hot and cold contacts and evaluate the state closure residual $R_f$.

Figure~\ref{fig:separation}(a) shows that changing $L$ at fixed controls generally violates the state closure condition. Its two curves correspond to reference protocols with different $a_{B2}$ values; as specified in~\ref{app:numericalprocedure}. It is observed that, for different $a_{B2}$, the detector separation must be adjusted to enforce state closure. This further shows that, for the full set of parameters defining a closed-cycle operating point (which satisfies all cycle conditions), any variation in one parameter necessitates compensatory changes in the others. Specifically, the detector separation does not enter the local response directly, but it modifies the nonlocal coherence constraints required for a closed cycle.

In figure~\ref{fig:separation}(b), the control parameters and initial populations are retuned at each separation to restore state closure and energy balance. Both the absorbed heat and gap-work efficiency then vary along the resulting closed-cycle branch. Each plotted point satisfies Eq.~\eqref{eq:closure}, the energy-balance condition, the required heat and work signs, and the positivity checks. This dependence is indirect: it arises because changing $L$ alters the coherence-closure constraints Eq.~\eqref{eq:cohclosure} and hence the local controls, rather than because separation enters the local population response. Together with the separable post-hot-stroke states identified in Sec.~\ref{sec:branch}, these results show that generated entanglement is not necessary for the enhancement mechanism considered here.

\section{Conclusions}\label{sec:conclusion}

We have investigated a two-detector Unruh quantum Otto engine whose working medium is initially prepared in an uncorrelated product state. We identified numerical operating points satisfying initial-state passivity, state closure, energy balance, and the requisite heat--work sign conditions. At these points, the efficiency exceeds the conventional Otto value, as quantified by Eq.~\eqref{eq:enhance}.

Our analysis distinguishes the roles of the diagonal populations and the off-diagonal coherences generated during detector--field interactions. Under the diagonal free Hamiltonian adopted here, the expectation value of the energy and the corresponding heat and work expressions depend only on the level populations. Generated coherence therefore provides no direct contribution to these energy quantities. In particular, enhancement occurs even when the post-hot-stroke state is separable. Thus entanglement generation is not required for the efficiency-enhancement mechanism identified within this model.

The mechanism is the interplay between unequal proper-time weighting and the local population changes. Expressing both detectors' energies with respect to detector $A$'s proper time introduces the acceleration ratio $\alpha_H$ into the heat contribution from detector $B$. In the equal-acceleration limit, the efficiency reduces to the conventional Otto value. When detector $A$ has the smaller acceleration, a positive population change of detector $B$ produces an enhanced efficiency ratio, provided that the absorbed heat is positive. An admissible engine cycle additionally requires state closure, energy balance, and positive work output.

Although the generated coherences do not directly alter the efficiency at fixed populations, they constrain whether the complete detector state can return to its initial value. For the product initial state and Gaussian switching functions considered here, the imaginary part of the exchange coherence leads to the same-sign phase exclusion in Eq.~\eqref{eq:nogo}. The coherent terms therefore restrict the cold-stroke parameters that can satisfy state closure. Retuning these parameters can indirectly affect the efficiency, even when the population-dependent energy expressions remain unchanged at fixed local controls.

The separation analysis illustrates this distinction. Changing the transverse detector spacing modifies the nonlocal coherence terms and can destroy closure of a previously admissible cycle; when the controls are retuned to restore closure, both the absorbed heat and the work efficiency can vary. These results separate the local population responses that govern the specified energy exchanges from the nonlocal coherence constraints that govern cyclic operation.

The present cycles involve two independent vacuum contacts and are closed only through the retained perturbative order. A complete operational model would require finite, continuously joined detector trajectories, a consistent treatment of the field state between contacts, and an accounting of switching, field-preparation, and motion-control costs. Such contributions are relevant to any claim of net work extraction from the vacuum \cite{Gray2018Scalar,Gallock2023Relativistic}. Controlling higher-order remainders is also necessary to establish exact finite-coupling separability and efficiency. Within these limitations, our results identify a local-population mechanism for enhancement of the UQOE efficiency and clarify the different thermodynamic roles of populations and nonlocal coherences.

\begin{acknowledgements}
	
	Pan and Cai acknowledge support from the National Natural Science Foundation of China (NSFC) (Grant Nos. 12605096 and 12547103), and the NSAF (Grant No. U2130205), and the Hainan Provincial Science and Technology Talent Innovation Project (No. KJRC2023B11). Zhang acknowledges support from the NSFC under Grant No. 12375057.
	
\end{acknowledgements}

\appendix

\section{Thermal response and state evolution}
\label{app:response}

\subsection{A phase restriction on state closure}
\label{sec:phase}

The complex-valued response matrices impose a state closure constraint in addition to energy balance. For stroke $j\in\{H,C\}$, define $x_{Dj}=\omega_j/a_{Dj}$, where $\omega_H=\omega_2$ and $\omega_C=\omega_1$, and introduce \cite{Hu2015Entanglement}
\begin{equation}
	\begin{aligned}
		\cosh\kappa_j&=\frac12\left(\frac{a_{Aj}}{a_{Bj}}+\frac{a_{Bj}}{a_{Aj}}+a_{Aj}a_{Bj}L^2\right),\\
		\Theta_j&=\frac{x_{Aj}+x_{Bj}}{2}\kappa_j.
	\end{aligned}
	\label{eq:kappa}
\end{equation}
The local excitation and deexcitation coefficients satisfy the exact identity
\begin{equation}\label{eq:commutator}
	V_{Dj}-U_{Dj}=\frac{s_jx_{Dj}}{2\sqrt\pi}.
\end{equation}
Cancellation of the diagonal population changes determines the initial excited-state probabilities
\begin{equation}\label{eq:pclosure}
	\begin{aligned}
		p_D&=\frac{U_{DH}+U_{DC}}{U_{DH}+V_{DH}+U_{DC}+V_{DC}}\\
		&=\frac{U_D^\Sigma}{2U_D^\Sigma+\sum_js_jx_{Dj}/(2\sqrt\pi)}.
	\end{aligned}
\end{equation}
The local excitation response is positive and decreases monotonically with the dimensionless gap $x_D$. As shown in \ref{app:spectrum}, it is a Gaussian convolution of the strictly decreasing function
$\nu/(e^{2\pi\nu}-1)$. Since $x_{Aj}>x_{Bj}>0$ in both strokes, $U_A^\Sigma<U_B^\Sigma$. Defining $c_D=\sum_j s_jx_{Dj}/(2\sqrt\pi)>0$ also gives $c_A>c_B$. For positive $U$ and $c$, the function $p(U,c)=U/(2U+c)$ satisfies
\begin{equation}
	\frac{\partial p}{\partial U}=\frac{c}{(2U+c)^2}>0,\quad
	\frac{\partial p}{\partial c}=-\frac{U}{(2U+c)^2}<0.
	\label{eq:pmonotonic}
\end{equation}
It follows that $p(U_A^\Sigma,c_A)<p(U_B^\Sigma,c_A) <p(U_B^\Sigma,c_B)$ and, consequently,
\begin{equation}
	0<p_A<p_B<\frac12.
	\label{eq:passive}
\end{equation}

Retaining the imaginary contribution of the time-ordered cross kernel yields
\begin{equation}
	\begin{aligned}
		&\operatorname{Im}Y_{23,j}=(p_B-p_A)A_j\cos\Theta_j,\\
		&A_j=\frac{s_j}{4\sqrt\pi\sinh\kappa_j}
		\exp\left[-\frac{s_j^2(x_{Aj}-x_{Bj})^2}{4} -\frac{\kappa_j^2}{4s_j^2}\right]>0.
	\end{aligned}\label{eq:imag}
\end{equation}
The common free-evolution phase does not rotate $Y_{23}$. Its imaginary closure condition is $(p_B-p_A) (A_H\cos\Theta_H+A_C\cos\Theta_C)=0$. For two distinct pointlike detectors and finite positive control parameters, this implies
\begin{equation}
	\cos\Theta_H\cos\Theta_C>0
	\quad\Longrightarrow\quad R_f\ne0.
	\label{eq:nogo}
\end{equation}
In particular, state closure is excluded when both phases lie in the first positive-cosine region, $0\leq\Theta_H,\Theta_C<\pi/2$. Opposite cosine signs remove this obstruction but are not sufficient for closure: the real part of the exchange-coherence condition and the complex $Y_{14}$ condition must also be satisfied.

The restriction in Eq.~\eqref{eq:nogo} relies on Gaussian switching, equal detector--field coupling strengths in both contacts, the acceleration ratios defined in Eq.~\eqref{eq:alphas}, and the absence of an independent relative local phase operation. Different switching profiles, unequal coupling strengths, or additional phase controls may change the closure constraints and require separate analysis. 

\subsection{Reduced-state correction and its matrix entries}

We derive the state correction generated by a single detector--field interaction. Stroke labels are suppressed below; the expressions apply separately to the hot and cold contacts with their respective parameters. In the ordered basis $\{|gg\rangle,|eg\rangle,|ge\rangle,|ee\rangle\}$, the initial state is given by Eq.~\eqref{eq:rho0}. It is an uncorrelated product state with neither initial coherence nor entanglement.

After one interaction stroke, the state takes the form
\begin{equation}
	\rho_{\mathrm{out}} = \rho_0+\lambda^2R[\rho_0]+\ord(\lambda^4),
\end{equation}
with
\begin{equation}
	R[\rho_0]=\begin{pmatrix}
		S_1&0&0&Y_{14}\\
		0&S_2&Y_{23}&0\\
		0&Y_{23}^*&S_3&0\\
		Y_{14}^*&0&0&S_4
	\end{pmatrix},
	\label{eq:R}
\end{equation}
where
\begin{align}
	S_1&=q_2V_A+q_3V_B-q_1(U_A+U_B),\nonumber\\
	S_2&=q_1U_A+q_4V_B-q_2(V_A+U_B),\nonumber\\
	S_3&=q_1U_B+q_4V_A-q_3(U_A+V_B),\nonumber\\
	S_4&=q_2U_B+q_3U_A-q_4(V_A+V_B),\label{eq:S}\\
	Y_{14}&=q_2G_{++}^{-}+q_3G_{++}^{+}-q_4F_{++}-q_1F_{++}^*,\nonumber\\
	Y_{23}&=q_1G_{+-}^{+}+q_4G_{+-}^{-}-q_3F_{+-}-q_2F_{+-}^*.
	\label{eq:Y}
\end{align}
The correction matrix is Hermitian and traceless. In particular, $S_2+S_4=(1-p_A)U_A-p_AV_A$ and $S_3+S_4=(1-p_B)U_B-p_BV_B$, giving the local population-response expressions in Eq.~\eqref{eq:localpop} at the coefficient level. Taking the imaginary part of the final expression in Eq.~\eqref{eq:Y} yields $(q_3-q_2)I_s=(p_B-p_A)I_s$, consistent with Eq.~\eqref{eq:imag}. All coefficients in $R[\rho_0]$ exclude the overall factor $\lambda^2$. The real coefficients $S_1,S_2,S_3,S_4$ describe signed changes in the populations of $|gg\rangle$, $|eg\rangle$, $|ge\rangle$, and $|ee\rangle$, respectively. The complex coefficients $Y_{14}$ and $Y_{23}$ describe coherences between $|gg\rangle$ and $|ee\rangle$, and between $|eg\rangle$ and $|ge\rangle$, respectively.

To express the off-diagonal entries, define
\begin{equation}
	\begin{aligned}\label{eq:crossnotation}
		G_{++}^{\pm} &= G_s\left( x_A+x_B,\,\pm\frac{x_A-x_B}{2};\kappa \right),\\
		F_{++} &= F_s\left( x_A+x_B,\, \frac{x_A-x_B}{2};\kappa \right), \\
		G_{+-}^{\pm} &= G_s\left( x_A-x_B,\,\pm\frac{x_A+x_B}{2};\kappa \right),\\
		F_{+-} &= F_s\left( x_A-x_B,\, \frac{x_A+x_B}{2};\kappa \right), 
	\end{aligned}
\end{equation}
The functions $G_s$ and $F_s$, derived in~\ref{app:spectrum}, represent Gaussian-integrated unordered and time-ordered cross responses, respectively. The $G$ terms arise from first-order interactions acting on opposite sides of the density matrix, whereas the $F$ terms arise from second-order, time-ordered contributions acting on either the ket or bra. Their population-dependent prefactors are determined by the initial probabilities $q_i$.

The subscript $++$ labels the frequency combination associated with pair coherence between $|gg\rangle$ and $|ee\rangle$, while $+-$ labels coherence between the single-excitation states. The superscripts $\pm$ on $G$ select the sign of its second frequency argument and do not denote complex conjugation.


We evaluate the matrix elements in Eq.~\eqref{eq:R}. Writing the detector proper times in terms of the dimensionless boost times $u_D=a_D\tau_D$, we use a common Gaussian width $s$
\begin{equation}
	\chi_D(\tau_D) = \exp\left[-\frac{(a_D\tau_D)^2}{2s^2}\right].
\end{equation}
The corresponding proper-time width for detector $D$ is $\Delta=s/a_D$. Thus, equal widths $s$ do not imply equal proper-time switching widths for detectors with different accelerations.

\subsection{Response coefficients from the boost spectrum}
\label{app:spectrum}

We now derive the vacuum two-point function along the detector trajectories and its boost-frequency spectrum. For $u=u_A-u_B$ and $\kappa>0$, the pullback of the massless-field Wightman function, including the proper-time Jacobians, is
\begin{equation}
	\frac{W_{AB}(u)}{a_Aa_B} = -\frac{1}{8\pi^2} \frac{1}{\cosh(u-\ii0^+)-\cosh\kappa}.
	\label{eq:wightman}
\end{equation}
Here $W_{AB}$ denotes the vacuum two-point function evaluated on the worldlines of detectors $A$ and $B$. The factors $1/a_A$ and $1/a_B$ arise when the proper-time integrals are reparametrized using boost time. The dimensionless parameter $\kappa$ depends on the acceleration ratio and transverse separation $L$. The singularities at $u=\pm\kappa$ correspond to null-related events on the two worldlines.

The local correlation function is recovered in the distributional limit $\kappa\to0$. We adopt the Fourier convention
\begin{equation}
	\mathcal R_\kappa(\nu) = \int_{-\infty}^{\infty}\dd u\, e^{-\ii\nu u} \frac{W_{AB}(u)}{a_Aa_B},
\end{equation}
Contour integration over the poles $u=\pm\kappa+2\pi\ii n+\ii0^+$ yields
\begin{equation}
	\mathcal R_\kappa(\nu)=\frac{\sin(\nu\kappa)}{2\pi\sinh\kappa\,(e^{2\pi\nu}-1)},\quad
	\mathcal R_0(\nu)=\frac{\nu}{2\pi(e^{2\pi\nu}-1)}.
	\label{eq:spectrum}
\end{equation}
The frequency $\nu$ is conjugate to dimensionless boost time. At $\nu=0$, the spectral functions are defined by continuity
\begin{equation}
	\mathcal R_\kappa(0)=\frac{\kappa}{4\pi^2\sinh\kappa},
	\quad\mathcal R_0(0)=\frac{1}{4\pi^2}.
\end{equation}
Under this Fourier convention, positive detector energy gap $\omega$ corresponds to excitation.

The local spectrum $\mathcal R_0$ determines the response along an individual detector trajectory, whereas the cross spectrum $\mathcal R_\kappa$ encodes correlations between the two trajectories. Because it contains the oscillatory factor $\sin(\nu\kappa)$, the latter is not a nonnegative single-detector transition spectrum.

Introduce the mean and relative boost-time coordinates
\begin{equation}
	q=\frac{u_A+u_B}{2}, \quad u=u_A-u_B.
\end{equation}
The integration coordinate $q$ should not be confused with the initial probabilities $q_i$. The product of the switching functions then becomes
\begin{equation}
	e^{-u_A^2/(2s^2)}e^{-u_B^2/(2s^2)}
	=
	e^{-q^2/s^2-u^2/(4s^2)}.
\end{equation}
For real $k_+$ and $k_-$, the detector phase factor takes the form $e^{-\ii k_+q-\ii k_-u}$. Thus $k_+$ and $k_-$ are conjugate to the mean and relative boost times, respectively. The combinations relevant to pair and exchange coherences are listed in Eq.~\eqref{eq:crossnotation}.

The Gaussian integrations can be evaluated analytically \cite{pozas2015harvesting}
\begin{equation}
	\begin{aligned}
		\int_{-\infty}^{\infty}\dd q\,e^{-q^2/s^2-\ii k_+q}
		&=\sqrt{\pi}s\,e^{-s^2k_+^2/4},\\
		\int_{-\infty}^{\infty}\dd u\,
		e^{-u^2/(4s^2)-\ii(k_--\nu)u}&=2\sqrt{\pi}s\,e^{-s^2(k_--\nu)^2}.
	\end{aligned}
\end{equation}
Substituting the inverse Fourier transform of Eq.~\eqref{eq:spectrum} yields
\begin{equation}
	G_s(k_+,k_-;\kappa)=s^2e^{-s^2k_+^2/4}\int_{-\infty}^{\infty}\dd\nu\,
	\mathcal R_\kappa(\nu)e^{-s^2(k_--\nu)^2}.
	\label{eq:gain}
\end{equation}
The coefficient $G_s$ is a Gaussian-weighted average of the field spectrum over a boost-frequency interval of width of order $1/s$. 

The individual detector responses follow by taking the local limit $\kappa=0$ with the appropriate transition frequencies \cite{fewster2016waiting}
\begin{equation}
	U_D=G_s(0,x_D;0), \quad V_D=G_s(0,-x_D;0), \quad x_D=\omega/a_D. \label{eq:localUV}
\end{equation}
Both coefficients are dimensionless and nonnegative. They depend on the detector's acceleration, energy gap, and switching width.

The time-ordered cross-response coefficient is
\begin{equation}
	\begin{aligned}
		F_s(k_+,k_-;\kappa)=&\tfrac12\bigl[G_s(k_+,k_-;\kappa)+G_s(k_+,-k_-;\kappa)\bigr]\\&-\ii I_s(k_+,k_-;\kappa),\\
		I_s(k_+,k_-;\kappa)=&\frac{s}{4\sqrt{\pi}\sinh\kappa}e^{-s^2k_+^2/4-\kappa^2/(4s^2)}\cos(k_-\kappa). \label{eq:ordered}
	\end{aligned}
\end{equation}
The real contribution arises from the symmetrized field two-point function, while the imaginary contribution originates from the field commutator in the Dyson time-ordered term. For the massless field, the cross commutator has support at the null separations $u=\pm\kappa$. Evaluating the switching functions and detector phases at these points produces the exponential and cosine factors in $I_s$ \cite{tjoa2021entanglement,maeso2022entanglement}. The term $I_s$ is therefore a physical contribution to the complex response, not an artifact of numerical regularization. Discarding the imaginary part of $F_s$ would alter the state coherences and could lead to an incorrect state closure condition.

Since $G_s$ is real, Eq.~\eqref{eq:Y} gives
\begin{equation}
	\begin{aligned}
		\operatorname{Im}Y_{23} &= (q_3-q_2) I_s\left( x_A-x_B,\frac{x_A+x_B}{2};\kappa \right)\\ &= (p_B-p_A) I_s\left( x_A-x_B,\frac{x_A+x_B}{2};\kappa \right).
	\end{aligned}
\end{equation}
This relation establishes Eq.~\eqref{eq:imag} and shows that the imaginary exchange coherence is proportional to the initial population imbalance. In particular, it vanishes when $p_A=p_B$, even though the real part of the coherence may remain nonzero.

The cross responses also satisfy the identity
\begin{equation}
	\begin{aligned}
		G_s&(k_+,-k_-;\kappa)-G_s(k_+,k_-;\kappa) \\&= \frac{se^{-s^2k_+^2/4-\kappa^2/(4s^2)}}{2\sqrt{\pi}\sinh\kappa}
		\sin(k_-\kappa).
	\end{aligned}
	\label{eq:crossidentity}
\end{equation}
This follows from
\begin{equation}
	\mathcal R_\kappa(-\nu)-\mathcal R_\kappa(\nu)
	=
	\frac{\sin(\nu\kappa)}{2\pi\sinh\kappa}
\end{equation}
together with the Gaussian frequency integral. Setting $k_+=0$ and $k_-=x_D$ and then taking the local limit gives
\begin{equation}
	V_D-U_D = \frac{s x_D}{2\sqrt{\pi}}, \label{eq:localUVdifference}
\end{equation}
This reproduces Eq.~\eqref{eq:commutator} without the stroke label and holds for any finite Gaussian width; no long-interaction limit is required.

\subsection{Unruh thermal response and inertial limits}

To verify the consistency of the response functions used in our two-detector UQOE model, we examine their long-interaction-time and inertial limits. In the long-interaction-time limit, the local response recovers the standard Unruh transition rates and the corresponding detailed-balance relation. In the inertial limit, finite-time switching produces a nonvanishing excitation response and allows the generation of nonlocal coherence between the detectors. These limiting results provide independent checks of the local and cross-response functions entering the reduced density matrix, which are essential for calculating heat, work, and state cycle closure \cite{Hu2022Nonidentical}.

At fixed acceleration and energy gap, increasing $s$ broadens the interaction in proper time while narrowing its frequency window. In the long-interaction limit, the Gaussian approaches a delta distribution:
\begin{equation}
	\frac{s}{\sqrt{\pi}} e^{-s^2(\nu-x_D)^2} \longrightarrow \delta(\nu-x_D).
\end{equation}
Consequently
\begin{equation}
	\frac{U_D}{\sqrt{\pi}s} \longrightarrow \mathcal R_0(x_D),\qquad
	\frac{V_D}{\sqrt{\pi}s} \longrightarrow \mathcal R_0(-x_D).
\end{equation}

The transition rates per unit proper time are obtained by dividing the response coefficients by the effective Gaussian duration
\begin{equation}
	\int_{-\infty}^{\infty}\dd\tau_D\,\chi_D^2(\tau_D)=\frac{\sqrt{\pi}s}{a_D}.
\end{equation}
Using Eq.~\eqref{eq:spectrum}, we obtain the limiting excitation and deexcitation rates, omitting the common factor $\lambda^2$
\begin{equation}
	\begin{aligned}
		\frac{U_D}{\sqrt{\pi}s/a_D} &\longrightarrow \frac{\omega}{2\pi} \frac{1}{e^{2\pi\omega/a_D}-1}, \\
		\frac{V_D}{\sqrt{\pi}s/a_D} &\longrightarrow \frac{\omega}{2\pi} \frac{1}{1-e^{-2\pi\omega/a_D}} =
		\frac{\omega}{2\pi} \left[ 1+\frac{1}{e^{2\pi\omega/a_D}-1} \right].
		\label{eq:Unruhrates}
	\end{aligned}
\end{equation}
These are the standard Unruh transition rates at temperature $a_D/(2\pi)$, in units with $ \hbar =c= k_{\mathrm B} =1$. The excitation rate is proportional to the thermal occupation number, whereas the deexcitation rate includes both stimulated and spontaneous emission.

In this limit, the excitation-to-deexcitation ratio obeys detailed balance
\begin{equation}
	\frac{U_D}{V_D} \longrightarrow e^{-2\pi\omega/a_D}. \label{eq:Unruhdetailedbalance}
\end{equation}
The thermal character of the response is therefore established by detailed balance, not merely by a nonzero excitation probability.

At finite $s$, the frequency averaging in Eq.~\eqref{eq:gain} introduces switching-dependent corrections to $U_D$ and $V_D$. Their ratio therefore need not equal the Boltzmann factor in Eq.~\eqref{eq:Unruhdetailedbalance}; a finite interaction should not be identified with complete thermalization at the Unruh temperature. The finite-time perturbative description additionally requires sufficiently small transition probabilities \cite{Rodriguez2016Finite,sriramkumar1996finite,Stargen2026Finite,azizi2025time}.

\paragraph{Inertial local response.}

To obtain the inertial limit, let $a_A=a_B=a\to0$ while holding the proper Gaussian width $\Delta=s/a$ and energy gap $\omega$ fixed. Then $s\to0$ and $\omega/a\to\infty$, with their product fixed at $\omega\Delta$. The excitation coefficient reduces to
\begin{equation}
	U\longrightarrow \frac{ e^{-z^2} - \sqrt{\pi}z\,\operatorname{erfc}(z) }{4\pi}, \qquad z=\omega\Delta.
	\label{eq:inertialU}
\end{equation}
Here $\operatorname{erfc}$ denotes the complementary error function, and $z$ is the gap measured in units of the proper switching width.

The finite-time excitation is not a thermal Unruh response \cite{sriramkumar1996finite,Stargen2026Finite,azizi2025time}, since the proper acceleration and Unruh temperature both vanish. Instead, switching the interaction on and off produces frequency broadening and allows transient excitation of an inertial detector in the Minkowski vacuum. At fixed positive $\omega$, this excitation coefficient vanishes as $\Delta\to\infty$.

The corresponding deexcitation coefficient follows from Eq.~\eqref{eq:localUVdifference}
\begin{equation}
	V-U
	\longrightarrow
	\frac{z}{2\sqrt{\pi}}.
	\label{eq:inertialVminusU}
\end{equation}
Dividing by the effective proper-time duration $\sqrt{\pi}\Delta$ gives the rate difference $\omega/(2\pi)$. Thus, in the long-duration inertial limit, the excitation rate vanishes while the deexcitation rate approaches the spontaneous-emission contribution.

\paragraph{Inertial pair coherence.}

For two detectors initially in their ground states and separated by $L>0$, we have $q_1=1$ and $q_2=q_3=q_4=0$. Eq.~\eqref{eq:Y} then gives $Y_{14}=-F_{++}^*$. In the same inertial limit
\begin{equation}	\label{eq:inertialY14}
	Y_{14}\longrightarrow-\frac{\Delta}{4\sqrt{\pi}L}e^{-\omega^2\Delta^2-L^2/(4\Delta^2)}
	\left[\operatorname{erfi}\!\left(\frac{L}{2\Delta}\right)+\ii\right].
\end{equation}
Here $\operatorname{erfi}$ is the imaginary error function, which takes real values for real arguments. The term containing $\operatorname{erfi}$ contributes to the real part of $Y_{14}$, whereas the imaginary part arises from the time-ordered field commutator.

The ground-state preparation is used only to check the response formulas; the closed-cycle operating points instead use mixed product initial states. The expression shows that a field can generate coherence between $|gg\rangle$ and $|ee\rangle$ without direct detector--detector coupling. Entanglement, however, depends on the difference between nonlocal coherence and local excitation noise. For a ground-state input, the entanglement criterion is $|Y_{14}|-\sqrt{U_AU_B}$; nonzero $Y_{14}$ alone is insufficient.



\section{Numerical methods and consistency checks}\label{app:numericalprocedure}

\begin{table*}[t]
	\centering
	\caption{Independent checks over all 100 closed-family points.
		Response matrices exclude $\lambda^2$; coefficients use
		the reference acceleration $a^*$.}
	\label{tab:verification}
	\begin{tabular}{lr}
		\toprule
		Check & Observed bound\\
		\midrule
		$\max\|\boldsymbol f\|_2$, 512-point quadrature & $<8.1\times10^{-13}$\\
		$\max\|\boldsymbol f\|_2$, adaptive quadrature & $<8.1\times10^{-13}$\\
		Largest matrix-entry difference, 512 versus adaptive & $<7.1\times10^{-17}$\\
		$\max|\mathcal{B}|/(\lambda^2a^*)$ & $<7.4\times10^{-18}$\\
		Smallest eigenvalue of $\rho_H^{[2]}$ & $>6.46\times10^{-6}$\\
		Smallest eigenvalue of $(\rho_H^{[2]})^{T_B}$ & $>6.46\times10^{-6}$\\
		Largest local transition probability & $<6.41\times10^{-4}$\\
		Difference between the two enhancement expressions & $<2.4\times10^{-16}$\\
		\bottomrule
	\end{tabular}
\end{table*}

\paragraph{Response evaluation.}

We calculate the response matrices $R_H[\rho_0]$ and $R_C[\rho_0]$ given in~\ref{app:response}, without the overall factor $\lambda^2$. The integration over the mean time is performed analytically, leaving only one-dimensional frequency integrals. We also retain the imaginary part of the time-ordered response to calculate the coherence terms correctly.

For numerical calculations, we separate the vacuum and thermal contributions. The integration range for the vacuum part extends 12 Gaussian widths beyond the peak, while the thermal part is integrated up to $\nu=16$. For deexcitation, we shift the integration variable so that the Gaussian peak remains within the integration range. Finally, we check the numerical results using both fixed-order and adaptive integration methods.

\paragraph{Construction of the 100-point closed family.}

The fixed control parameters used for Fig.~\ref{fig:branch} are
\begin{equation}\label{eq:fixedcontrols}
	\begin{aligned}
		\alpha_H&=0.9931002043581229,\\ s_H&=s_C=0.4992820253054886,\\
		a^*L&=1.9602737671581.
	\end{aligned}
\end{equation}

We sample 100 equally spaced values of $a_{B2}/a^*$ over the interval from $0.9$ to $1.2$. At each point, the nonlinear solver uses three variables
\begin{equation}
	x_B=\frac{\omega_2}{a_{B2}},\qquad
	r=\frac{\omega_1}{\omega_2},\qquad
	\gamma=\frac{a_{B4}}{r a_{B2}}.
	\label{eq:branchvariables}
\end{equation}
For each trial, we set $\omega_2=a_{B2}x_B$, $\omega_1=r\omega_2$, $a_{B4}=\gamma r a_{B2}$, and $\alpha_C=1-(1-\alpha_H)/r$, thereby imposing $\mathcal{B}=0$. The remaining accelerations follow as $a_{A2}=\alpha_Ha_{B2}$ and $a_{A4}=\alpha_Ca_{B4}$. We then evaluate the local responses in both strokes and determine $p_A,p_B$ from Eq.~\eqref{eq:pclosure}. These populations define the input state in Eq.~\eqref{eq:rho0}, from which the full complex response matrices are constructed.

The solver imposes three independent real coherence-closure constraints through the normalized residual
\begin{equation}
	\begin{aligned}
		\boldsymbol f=\{&
		\frac{|Y_{14,H}|-|Y_{14,C}|}{|Y_{14,H}|+|Y_{14,C}|},
		\frac{\operatorname{Re}(Y_{23,H}+Y_{23,C})}{|Y_{23,H}|+|Y_{23,C}|},\\
		&\frac{\operatorname{Im}(Y_{23,H}+Y_{23,C})}{|Y_{23,H}|+|Y_{23,C}|}\}.
	\end{aligned}\label{eq:residual}
\end{equation}
The denominators remain nonzero throughout the reported branch. Equality of the two $Y_{14}$ magnitudes permits a choice of the common phase satisfying $e^{2\ii\vartheta}Y_{14,H}+Y_{14,C}=0$. The remaining constraints enforce cancellation of the real and imaginary parts of $Y_{23}$. We subsequently verify the complete $4\times4$ residual in Eq.~\eqref{eq:closure}, including the diagonal elements. 

The numerical procedure first resolves the reference operating point from a coarse initial guess and then continues toward larger and smaller values of $a_{B2}$. Previously converged points provide initial guesses only: both response matrices are recomputed at every objective-function evaluation. The solver varies $(x_B,r,\gamma)$ within $[0.1,20]\times[0.03,0.98]\times[0.02,0.98]$ using least squares, with termination tolerances of $10^{-12}$ and a maximum of 300 function evaluations per point. The working integrals are evaluated using 256-point Gauss--Legendre quadrature, and every candidate is independently recomputed with 512-point and adaptive quadrature.

Accepted points must satisfy $\|\boldsymbol f\|_2<10^{-7}$, positive $Q_H$ and $W_{\rm gap,out}$, negative $Q_C$, and an energy-balance residual below $10^{-12}$ in dimensionless coefficient units. We also verify $0<p_A<p_B<1/2$ and $0<\alpha_C<\alpha_H<1$. The independently recomputed hot- and cold-stroke matrix elements must agree within $10^{-10}$. The observed discrepancies are substantially smaller, as summarized in Table~\ref{tab:verification}. Along the accepted branch, $X_B>0$ and the efficiency enhancement is positive.

\paragraph{Reference point and numerical ranges.}

At $a_{B2}=a^*$, the reference operating point is
\begin{equation}
	\begin{aligned}
		\omega_2/a^*&=1.6927973534863,&r&=0.4738891826941,\\
		\gamma&=0.4753613252479,&p_B&=0.0175690384783,\\
		\alpha_C&=0.9854400651168,&p_A&=0.0170379319315.
		\label{eq:reference}
	\end{aligned}
\end{equation}
The required free-evolution phase is $2\vartheta=-2.163589619318$ modulo $2\pi$. The exchange phases, $\Theta_H=2.9381745445$ and $\Theta_C=1.5611038104$, have cosines of opposite signs. The corresponding energies are
\begin{equation}
	\begin{aligned}
		\frac{Q_H}{\lambda^2a^*}&=0.0278214992110,
		\frac{Q_C}{\lambda^2a^*}=-0.0131333674564,\\
		\frac{W_{\rm gap,out}}{\lambda^2a^*}&=0.0146881317547.
		\label{eq:referenceenergies}
\end{aligned}\end{equation}
For this point, $\eta_{\rm gap}=0.5279417778$, $\eta_0=0.5261108173$, and $X_B/\lambda^2=0.0082897431$. Over the complete 100-point family,
\begin{align}
	8.75929\times10^{-4}&\leq\Delta\eta_{\rm gap}
	\leq2.64033\times10^{-3},\nonumber\\
	1.54885\times10^{-3}&\leq X_B/\lambda^2
	\leq1.60086\times10^{-2},\nonumber\\
	3.46171\times10^{-3}&\leq\zeta\leq3.48337\times10^{-3}.
	\label{eq:gainrange}
\end{align}
Here $\Delta\eta_{\rm gap}$ is the efficiency difference. The branch spans $0.97179<\alpha_C<0.99078$, with $p_A$ ranging approximately from $0.00246$ to $0.03799$ and $p_B$ from $0.00261$ to $0.03872$. The efficiency ranges approximately from $0.2530$ to $0.7580$. Because $\zeta$ is nearly constant, the variation in $\Delta\eta_{\rm gap}=\eta_0\zeta/(1-\zeta)$ is largely controlled by the gap ratio.

\paragraph{State tests and processing of the branch data.}

At each point we construct the state $\rho_H^{[2]}=\rho_0+\lambda^2R_H[\rho_0]$ at $\lambda^2=10^{-3}$. We check Hermiticity, trace normalization, and the eigenvalues of both $\rho_H^{[2]}$ and its partial transpose. After verifying positivity, we evaluate $C_H^{[2]}$ using both branches of the X-state concurrence expression applied to Eq.~\eqref{eq:hotstatematrix}. All 100 truncated states are positive and satisfy the PPT criterion, yielding $C_H^{[2]}=0$. We also compute $K=\|(R_H[\rho_0])^{T_B}\|_{\rm op}$ and obtain
\begin{equation}
	\min_{\rm branch}(m_0-\lambda^2K)=3.19\times10^{-6}>0.
	\label{eq:numericalmargin}
\end{equation}
This value gives the maximum allowed norm of the unknown higher-order remainder for the sufficient separability condition in Eq.~\eqref{eq:PPTbound} to hold. It does not estimate the actual remainder.

\paragraph{Fixed-control and retuned separation scans.}

We begin with the reference closed-cycle operating point specified in Eq.~\eqref{eq:reference}, for which $a^{*}L_{\rm ref}=1.9602737671581$, $\alpha_H=0.9931002043581229$, and $s_H=s_C=0.4992820253054886$. For comparison, a second reference protocol is obtained by multiplying all accelerations and energy gaps by $1.1$ and dividing the reference separation by $1.1$, while keeping the Gaussian switching widths and initial populations unchanged. For Fig.~\ref{fig:separation}(a), we sample 100 separations $L$ in $[0.7L_{\rm ref},1.3L_{\rm ref}]$, choosing grids that include the reference separation of each protocol. At every separation, we recompute the complete complex response matrices $R_H[\rho_0](L)$ and $R_C[\rho_0](L)$ and evaluate the normalized closure residual
\begin{equation}
	\mathcal{R}_L =
	\frac{
		\left\|U_{\vartheta_{\rm ref}}R_H[\rho_0](L)
		U_{\vartheta_{\rm ref}}^\dagger+R_C[\rho_0](L)\right\|_F}{\left\|R_H[\rho_0](L)\right\|_F+
		\left\|R_C[\rho_0](L)\right\|_F}.
	\label{eq:fixedLresidual}
\end{equation}
Here $\vartheta_{\rm ref}$ is the free phase of the corresponding reference cycle. The residual measures the failure of state closure, including both population and coherence contributions. It vanishes at the reference separation of each protocol; nonzero values away from that point quantify the loss of closure at fixed controls. 

For Fig.~\ref{fig:separation}(b), we construct a distinct branch of closed-cycle operating points at 100 separations in $[0.9L_{\rm ref},1.3L_{\rm ref}]$, including $L_{\rm ref}$. The lower bound is selected because the broader exploratory range did not yield a numerically verified closed cycle at every sampled separation. Failure to locate a solution numerically does not establish its nonexistence.
At each $L$, we fix $a_{B2}$, $\alpha_H$, and the Gaussian switching widths and retune $\omega_2$, $r=\omega_1/\omega_2$, and $\gamma=a_{B4}/(r a_{B2})$. We impose
\begin{equation}
	\alpha_C=1-\frac{1-\alpha_H}{r},\quad
	a_{B4}=r\gamma a_{B2},\quad \omega_1=r\omega_2,
\end{equation}
which enforces the energy-balance relation. The initial populations are recalculated from the diagonal closure conditions, and the free phase is chosen to cancel the $Y_{14}$ coherence. The least-squares search then minimizes the three normalized real coherence residuals in Eq.~\eqref{eq:residual}.

\end{document}